\documentclass[prl,twocolumn,superscriptaddress,showpacs,epsf]{revtex4}
\usepackage{rotating}
\usepackage{graphicx}
\usepackage{latexsym}
\usepackage{amssymb}
\usepackage{amsfonts}
\usepackage{soul}
\usepackage{color}
\usepackage{amsmath}
\begin{document}

\title{Local mapping of dissipative vortex motion}

\author{B. Raes}
\affiliation{INPAC -- Institute for Nanoscale Physics and Chemistry, Nanoscale Superconductivity \\ and Magnetism Group, K.U.Leuven, Celestijnenlaan 200D, B--3001 Leuven, Belgium}
\author{J. Van de Vondel}
\affiliation{INPAC -- Institute for Nanoscale Physics and Chemistry, Nanoscale Superconductivity \\ and Magnetism Group, K.U.Leuven, Celestijnenlaan 200D, B--3001 Leuven, Belgium}
\author{A.V. Silhanek}
\affiliation{D\'{e}partement de Physique, Universit\'{e} de Li\`{e}ge, All\'{e}e du 6 ao\^{u}t 17, B5, B--4000 Sart Tilman, Belgium}
\author{C. de Souza Silva}
\affiliation{Departamento de Fisica, Universidade Federal de Pernambuco,
Cidade Universitaria, 50670-901 Recife-PE, Brazil}
\author{J. Gutierrez}
\affiliation{INPAC -- Institute for Nanoscale Physics and Chemistry, Nanoscale Superconductivity \\ and Magnetism Group, K.U.Leuven, Celestijnenlaan 200D, B--3001 Leuven, Belgium}
\author{R.B.G. Kramer}
\affiliation{Institut N\'{e}el, CNRS, Universit\'{e} Joseph Fourier, BP 166, 38042 Grenoble Cedex 9, France}
\author{V.V. Moshchalkov}
\affiliation{INPAC -- Institute for Nanoscale Physics and Chemistry, Nanoscale Superconductivity \\ and Magnetism Group, K.U.Leuven, Celestijnenlaan 200D, B--3001 Leuven, Belgium}

\date{\today}
\begin{abstract}
We explore, with unprecedented single vortex resolution, the
dissipation and motion of vortices in a superconducting ribbon under
the influence of an external alternating magnetic field. This is
achieved by combing the phase sensitive character of ac-susceptibility,
allowing to distinguish between the inductive-and dissipative
response, with the local power of scanning Hall probe microscopy. Whereas the
induced reversible screening currents contribute only inductively,
the vortices do leave a fingerprint in the out-of-phase component.
The observed large phase-lag demonstrates the dissipation of
vortices at timescales comparable to the period of the driving force
(i.e. 13 ms). These results indicate the presence of slow
microscopic loss mechanisms mediated by thermally activated hopping transport of vortices between
metastable states.
\end{abstract}

\pacs{74.78.-w 74.25.F- 74.25.Wx 74.40.Gh}

\maketitle
The universal problem of energy dissipation embodies the
irreversible conversion of work into heat in a dynamic system. This is
fundamentally connected with the inequality present in the second
law of thermodynamics,  expressing a monotonic increase of the
entropy or the fact that a perpetual motion machine of the second
kind does not exist\cite{Feynman}. The macroscopic quantum effect of
superconductivity is the closest thing to the idiom of a perpetual
motion machine, due to the presence of an energy gap between the sea
of condensed Cooper pairs and the fermionic quasi-particles, electrical current can circulate forever in a superconducting ring\cite{Tinkham}.
However, when a type II superconductor is in the mixed state, the motion of
vortices, perpendicular to supercurrents, results in a resistive
voltage drop and the unique hallmark of dissipationless transport
collapses\cite{Josephson}.
\newline
In general, whenever a dissipative system is subjected to a periodic
excitation, e.g. a crystal to electromagnetic radiation or a driven
damped harmonic oscillator, the periodic force will perform work to
drive the system through subsequent dissipative cycles. The
dissipative or frictional component of the system, related to a
non-conservative force, will induce a phase shift between the
response and the external drive. For example, the imaginary part of
the relative permittivity is closely related to the absorption
coefficient of a material\cite{Fox} or a phase-lag appears in the
motion of a damped harmonic oscillator. This close connection
between dissipation of energy and the out-of-phase component of the
system's response is used in spectroscopic measurements to gain information
concerning the nature and efficiency of the dissipation processes and
is in a one-to-one relationship with the system's equilibrium fluctuations through the fluctuation dissipation theorem\cite{Callen}.
\newline
To investigate the dissipative nature of electrical transport, when a
superconductor is in the mixed state, the integrated response of the
superconductor upon the application of an external alternating
magnetic field is recorded in measurements of the global
ac-susceptibility\cite{Gomory}. The action of the alternating magnetic field will
result in a complex response arising from a collection of
two contributors: the screening currents and the vortices.
\newline
\begin{figure*}[ht!]
\includegraphics*[width=\linewidth]{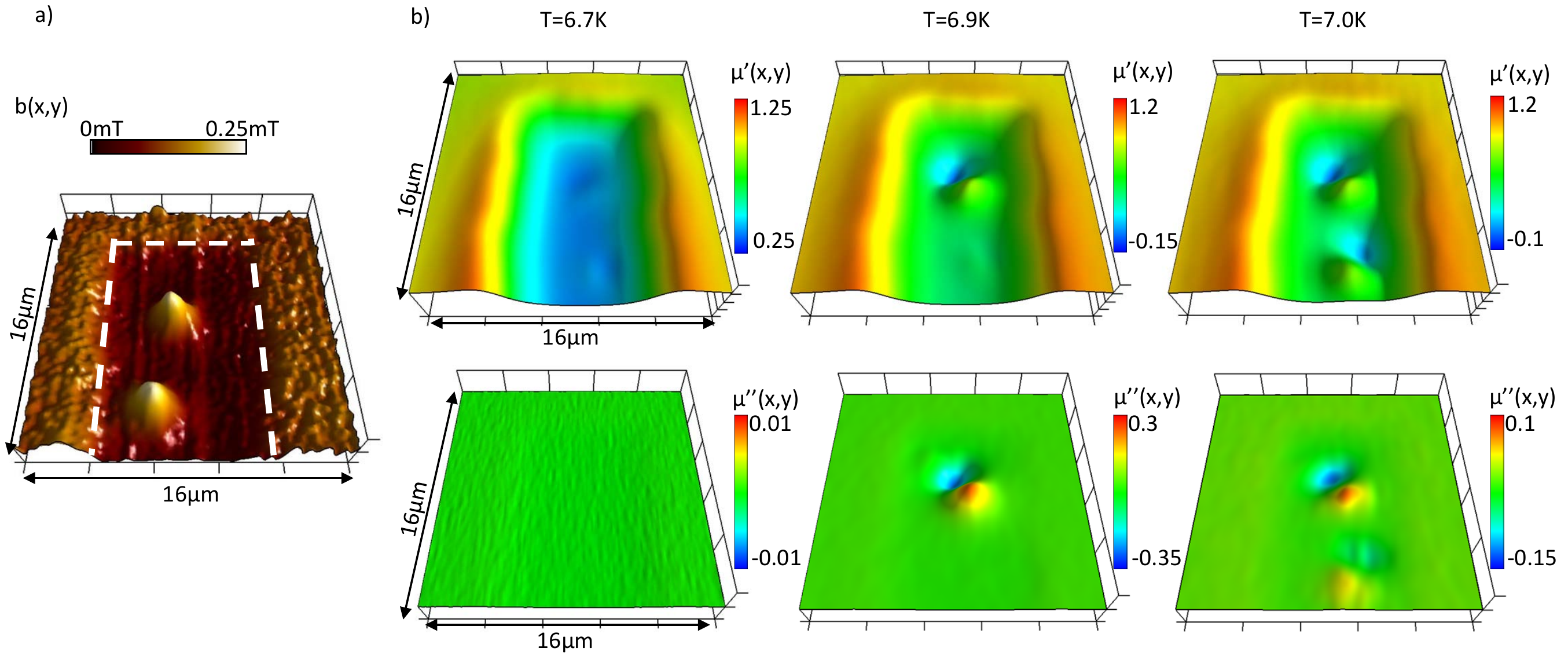}
\caption[]{(Color online) (a) Scanning Hall Probe Microscopy image of the local induction, b$_z$(x,y), acquired during shaking with an external applied ac field of amplitude, h$_{ac}$$=$0.1mT, and with frequency, f$=$77.123Hz at a temperature of T$=6.7$K. The initial vortex distribution is obtained by performing a field cool in an external applied dc magnetic field, H$=0.13$ mT. The white dashed line indicates the border of the Pb ribbon.(b) Simultaneously acquired maps of the real part of the relative permeability, $\mu'_r$(top row) and the imaginary part of the relative permeability, $\mu''_r$(bottom row), for different temperatures:(left to right) T$=$6.7K, 6.9K and 7.0K. }
\label{fig:1}
\end{figure*}
In first approximation one can study only the linear response, which
is completely determined by a measurement of the real and imaginary
part of the complex relative permeability. These
Fourier components are mutually connected by the Kramers-Kr\"onig
relationships. The real part describes the in-phase response of the
magnetic induction to the external magnetic ac field and is related to the
macroscopic shielding abilities. The imaginary part describes the out-of-phase response of the
magnetic induction, arising as indicated before, necessarily from
dissipative ac-losses within the superconductor. The dependencies of
these two response functions upon changing thermodynamical variables,
such as temperature and dc external magnetic field, or the
ac-excitation parameters such as the amplitude and the frequency,
provide very valuable information concerning the pinning efficiency and
reveal the fingerprints of the particular ac-dynamic phases the
vortex lattice exhibits\cite{Marchevsky, Koshelev, Pasquini}. Since the
recorded signal represents an average over all present flux lines
and screening currents in the sample, the link with the microscopic
ac-reponse is indirect. Pioneering theoretical works\cite{Campbell,
vanderbeek, Brandt} contributed substantially to link this global response
to the microscopic vortex dynamics and/or the ac-field penetration.
\newline
However, up till now the dissipation of individual vortices remains concealed,
despite the enormous amount of theoretical works. In this work we combine the strength of phase sensitive detection used in global ac-susceptibility measurements\cite{Gomory} and the power of
individual vortex visualization, accessible with Scanning Hall probe
microscopy (SHPM)\cite{Kirtley, DeFeo, Kramer, Kramer2}, to reveal the microscopic linear
response of a type-II superconductor to an external applied
ac-magnetic field. The local character of this scanning ac-susceptibility microscopy(SSM) technique and the argument
``Seeing is believing'' allow us to bridge the gap between the global ac-susceptibility measurements and their
associated microscopic theories of vortex motion and ac-field
penetration and the real microscopic ac-response.
\newline
\newline
In SSM, we continuously excite the sample with an external ac magnetic
field, h$_{ac}(t)$$=$$h_{ac}\cos(\omega t)$. The Hall voltage,
V$_H(x,y,t)$, measured locally by a Hall microprobe is picked up by
a lock-in amplifier. The excitation signal for the external applied
ac field, feeds a phase-locked loop which extracts the in-phase,
V$_1'(x,y)$, and quadrature components, V$_1''(x,y)$, of
V$_H(x,y,t)$. In the first approximation these are, respectively,
proportional to the in-phase, b$'_{ac}(x,y)$ and out-of-phase,
b$''_{ac}(x,y)$, ac-components of the local magnetic induction,
b$_z(x,y,t)$, coarse grained by the size of the cross. We can
introduce two response parameters which completely determine the
local linear response of b$_z(x,y,t)$ to h$_{ac}(t)$. We define the
real and imaginary part of the local relative permeability, $\mu_r(x,y)$, as\cite{Morozov}:
\begin{equation}\label{Eq0}
\begin{split}
\mu_r(x,y)&=\mu'_r(x,y)-i\mu''_r(x,y)\\
&=\frac{1}{h_{ac}}(b'_{ac}(x,y)-i b''_{ac}(x,y))
\end{split}
\end{equation}
Just as in the global ac-susceptibility technique, the in-phase component, $\mu'_r(x,y)$,
is related to the local inductive response, while the out-of-phase component, $\mu''_r(x,y)$, is
related to microscopic ac-losses. SSM provides a spatial map
of these two Fourier components and describes, as such, the local
linear response of b$_z(x,y,t)$ to $h_{ac}$(t) with single vortex
resolution.
\newline
The mapping of b$_z(x,y)$  was obtained using a modified
low-temperature SHPM from Nanomagnetics Instruments. The typical
scan area at 4.2K is 16$\times$16$\mu$m$^2$. The collinear DC and AC
external magnetic fields are always applied perpendicular to the
sample surface. The investigated sample is a Pb ribbon of 0.5 mm
long, 50 nm thick  and 9 $\mu$m wide and
exhibits a superconducting transition at T$_c$$=$7.20 K. As the
signal picked up by the Hall probe contains different contributions,
arising from the screening currents, the vortex signals and the
external field itself, the measured local linear ac-response is also
determined by all contributing factors. This particular sample design allows us to map
the spatial dependence of the linear response to h$_{ac}$(t), covering the whole width of the sample in a single scanning area, including the Meissner response at the sample border and the vortex motion deeper into the ribbon volume.
\newline
\newline
Fig.\ref{fig:1}a shows a SHPM image of a vortex distribution obtained after performing a field cool (FC) in H$=$0.13mT to T$=$6.7K while an external field with h$_{ac}$$=$0.1mT and f$=$77.123Hz, is continuously applied. The resulting vortex distribution obtained by performing a FC experiment, corresponds to a frozen vortex structure nucleated close to T$_c$\cite{Marchevsky1}. The FC process forces vortices to nucleate at the strongest pinning sites and results in a non-symmetrical vortex distribution. The external ac-field shows up as an additional monochromatic noise in the SHPM images getting more pronounced for temperatures close to T$_c$. However, for all investigated temperatures the average vortex positions do not change, indicating that for h$_{ac}$$=$0.1mT the resulting average vortex response is limited to displacements below the experimental spatial resolution.
\newline
Fig.\ref{fig:1}b shows a representative set of simultaneously acquired  SSM images of $\mu'_r(x,y)$ (top row) and $\mu''_r(x,y)$ (bottom row), respectively describing the inductive and dissipative response, when the temperature is decreased progressively from T$=$7K to T$=$6.7K. A first straightforward observation is that at the edges of the scan area, meaning relatively far away from the Pb ribbon, the local induction oscillates perfectly in-phase with h$_{ac}(t)$. A clear paramagnetic response, $\mu'_r(x,y)$$>$$1$, is visible at the edge of the Pb ribbon, where the response is dominated by the induced screening currents. This enhancement of the external ac-field is caused by a strong demagnetizing effect resulting from the thin film sample geometry\cite{Brandt1}. Upon entering the volume of the ribbon, we observe an increasing diamagnetic response as h$_{ac}(t)$ gets shielded by the screening currents. At the center of the Pb ribbon, a maximum diamagnetic response due to the screening current of $\mu'_r(x,y)$$=$0.27 at T$=$6.7K is reached, indicating an incomplete field expulsion. An important observation in Fig.\ref{fig:1}b is that the shielding currents do not show any contributing signal in $\mu''_r(x,y)$ for all temperatures, indicating that they are, within our experimental resolution, perfectly in-phase with the ac-excitation and as such are non-dissipative.
\newline
Within the ribbon volume the induced screening currents, $\mathbf{j}(t)$, will periodically shake the vortices, with a force: $\mathbf{f_L}(t)=\mathbf{j}(t)\times${\boldmath$\phi_0$}. The ac-dynamics of the vortices will crucially depend on the thermodynamical parameters of the SC system and the properties of the drive. For the described experimental parameters, the fingerprint of their motion in the SSM images, consists of two distinct unidirectional spots of opposite polarity surrounding the equilibrium vortex position. The inductive response can be easily interpreted. An area with a response exceeding the ac-response of the screening currents, $\mu'_r(x,y)$$>$$\mu'_r(x,y)_s$,  corresponds with a vortex, carrying an intrinsic positive local induction, moving in-phase with h$_{ac}(t)$ within this area. A region with a response lower than the ac-response of the screening currents, $\mu'_r(x,y)$$<$$\mu'_r(x,y)_s$, in some cases resulting even in a local negative permeability, $\mu'_r(x,y)$$<$$0$, indicates that b$_z(x,y,t)$ increases (decreases) upon decreasing (increasing) instantaneous h$_{ac}(t)$, corresponding with a vortex moving in anti-phase with h$_{ac}(t)$ within this area. A similar unique local negative $\mu'_r(x,y)$ response, but on a substantially larger spatial scale, has already been observed in the ac-dynamics of flux droplets in the presence of a geometrical barrier\cite{Morozov}. From thermodynamical considerations an overall integrated response between zero and one is expected for $<$$\mu'_r$$>$. Note however, that the meaning of the complex permeability as a macroscopic thermodynamical variable is lost in this local limit. Upon integrating the local signal over the whole scan area the expected non-negative response for $<$$\mu'_r$$>$ and $<$$\mu''_r$$>$ is recovered. This connection between $<$$\mu_r$$>$ as the integrand of the `local'  permeability, $\mu_r(x,y)$, which is directly related to the microscopic vortex dynamics, is used in theoretical models to explain the fingerprints of different dynamical VL regimes in measurements of the global ac-susceptibility and can be studied now directly by SSM. In sharp contrast to the screening currents' response, the vortices do leave a fingerprint in $\mu''_r(x,y)$ for sufficiently high temperatures. As such, the harmonic approximation of their motion exhibits an out-of-phase component. This indicates that vortex motion is accompanied by a dissipative process. The irreversible response disappears below T$<$6.8K, here the ac-response of the vortices is weak and, within the experimental resolution, perfectly in-phase. The particular depth and shape of the local pinning potential each vortex experiences has a profound effect on the ac-dynamics, i.e. at T$=$6.9K only one of the two vortices present in our scan area is shaken by h$_{ac}$. The nature of the microscopic processes resulting in vortex dissipation, each having a characteristic time, is a question of interest\cite{Kim,Bardeen,Tinkham1}.
\newline
\newline
\begin{figure*}[ht!]
\includegraphics*[width=\linewidth]{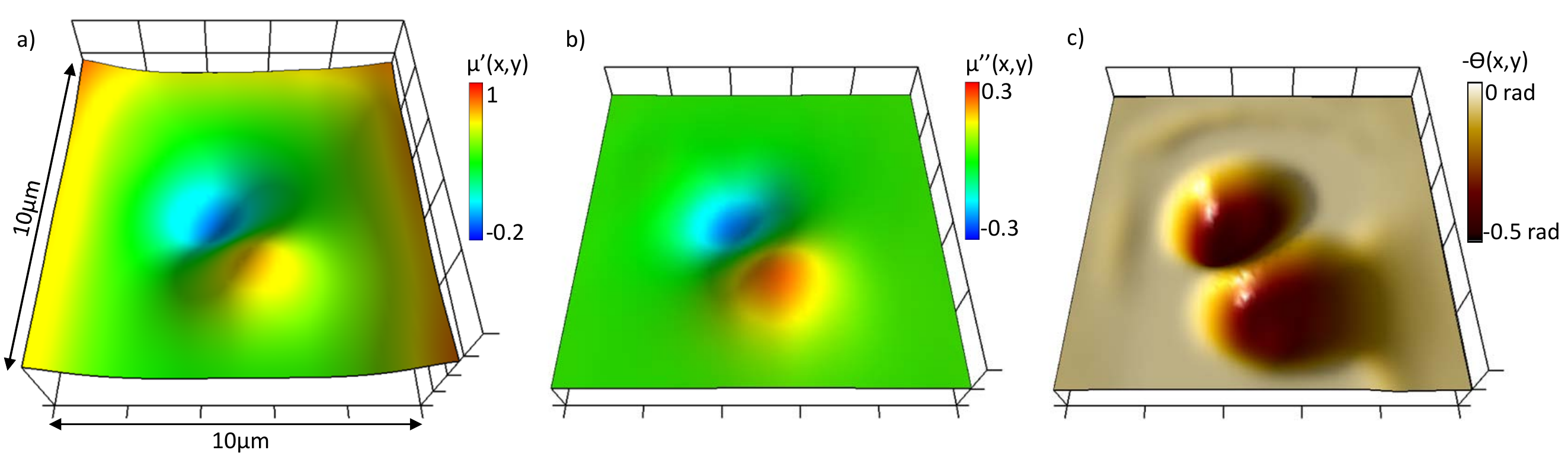}
\caption[]{(Color online) (a) Scanning Susceptibility Microscopy image of the real part of the
relative permeability, $\mu'_r$ for a single vortex upon shaking
with an external ac magnetic field of amplitude, h$_{ac}$$=$0.1mT, and frequency f$=$77.123Hz at a temperature of T$=6.9$K. The initial
vortex distribution is obtained by performing a field cool in an external applied dc magnetic field, H$=0.13$
mT.(b) Simultaneously acquired map of the imaginary part of the
relative permeability, $\mu''_r$. (c) Calculated spatial dependence
of minus the phase angle.} \label{fig:2}
\end{figure*}
In general, vortex-dynamics can be described by a phenomenological force balance equation of the form\cite{Gittleman}:
\begin{eqnarray}\label{Eq1}
m \mathbf{\ddot{x}}(t)-\mathbf{f_L}(t)+ \nabla U= -\eta \mathbf{\dot{x}}(t)+\boldsymbol\xi(t)
\end{eqnarray}
Where $\mathbf{x}(t)$ is the vortex position.
Here the left hand side describes the conservative part of the
dynamics, including a combination of an inertial term, the periodic
Lorentz drive, $\mathbf{f_L}(t)$ and the local potential, U, the vortex experiences due to a
combination of interactions with other vortices, the surface and the
local quenched disorder. For small excitations the local potential
in the equation of motion can be approximated by a harmonic
potential with spring constant $\alpha_L$, called the Labusch
constant\cite{Labusch}. The inertial term, including the mass per
unit length of a vortex, m, is pure effective in nature, as
a vortex is unable to sustain its existence outside the condensate
medium. It is accepted to be very small\cite{Suhl,Golubchik} so that there is
a short initial period of acceleration needed to reach the steady
state motion we consider. The right hand side of eq.\ref{Eq1}, describes the
effects of the non-conservative environment presenting energy
dissipation and modeled as a combination of a viscous friction,
$\eta$, and a random thermal force, $\boldsymbol\xi(t)$\cite{Fulde}. It
is this right hand side of the equation of motion which describes the
coupling to the environment. The linear
approximation to the steady state solution of eq.\ref{Eq1}, has
the following general form\cite{vanderbeek, Coffey}:
\begin{eqnarray}\label{Eq2}
\mathbf{x(t)}=-(\frac{\alpha_L}{1-i/\omega \tau_1}+i\omega
\eta)^{-1} \mathbf{f_L}(t)
\end{eqnarray}
Here $\tau_1=(\eta/\alpha_L)I_0^2(U(j)/2k_BT)$ is a characteristic
relaxation time related to thermally activated hopping of the
vortex, I$_0$(x) is the modified Bessel function and U(j) describes
an effective activation energy which is a combination of the intrinsic
pinning potential energy, U(j$=$0) and the Lorentz force energy
resulting from the induced super currents, $\mathbf{j}(t)$.
\newline
The solution, given by eq.\ref{Eq2}, directly shows  the
out-of-phase component in the linear response induced by the
aforementioned dissipative mechanisms. The term $i\omega\eta$
represents the viscous damping mechanism. It is
connected by an elementary model developed by Bardeen and
Stephen\cite{Bardeen} with resistive processes in the normal core
and by a more rigorous analysis with a finite intrinsic relaxation
time of the SC order parameter\cite{Tinkham1}. This dissipative
process has a typical short characteristic time of the order of,
$\tau_p$$=$$\eta/\alpha_L$$\leq$0.1$\mu$s\cite{Gittleman, Awad}. For
the applied low driving frequency, f$=$77.123 Hz, the
restoring force dominates over the viscous drag force, as
$\omega$$<<$1/$\tau_p$ and this term can be neglected. The term
$i/\omega \tau_1$ is related to thermally activated vortex hopping
across an effective activation barrier, following the classic
ideas of Anderson and Kim\cite{Anderson} and results from
$\boldsymbol\xi(t)$ in eq.\ref{Eq1}. This activated dissipation
process is typically associated with longer characteristic time
scales\cite{Hanggi}. Under certain conditions it is expected to
contribute substantially in our low frequency SSM experiment.
\newline
At low temperatures, when U(j)$>>$k$_B$T and thermally activated
flux motion can be neglected, $\tau_1$ diverges exponentially and
the character of the ac-response, $\mathbf{x(t)}=-\alpha_L
\mathbf{f_L}(t)$, is a pure reversible harmonic
motion as described by Campbell and Evetts\cite{Campbell}. This behaviour
explains the absence of a response in the SSM images of
$\mu''_r(x,y)$ for T$<$6.8K, while a response is still visible in
$\mu'_r(x,y)$. As the temperature rises, the thermal activation
energy decreases and $1/\omega\tau_1$ becomes appreciable, meaning
thermally activated vortex jumps between metastable states come into
play and contribute substantially to the dissipation process. This
explains the observed out-of-phase component for T$>$6.8K.
Fig.\ref{fig:2} shows a zoom on the ac-response of a single vortex
for T$=$6.9K and the corresponding spatial dependence of the
calculated phase, where we use a cutoff for $\mid\mu'_r(x,y)\mid$$<$$0.15$
to limit the divergence of arctan and we subtracted the contribution
of the screening currents in $\mu'_r(x,y)$. As shown in Fig.2(c),
the obtained phase shift is $\Theta$$=$-0.5rad. From eq.2, the phase
shift between the response and the drive is given by
$\Theta$$=$$-\arctan(1/\tau_1\omega)$. As $\tau_p$$\leq$0.1$\mu$s,
we obtain a lower limit for the effective activation barrier height
of U(j)$\geq$8.50$\times$10$^{-3}$eV$\sim$14.3 k$_B$T, similar to
typical average effective barrier heights found in the literature by macroscopic measurements\cite{Lange}.
\newline
The temperature dependence of the phase shift shows a maximum at
T$=$6.85K. Optimal energy dissipation is expected when the driving frequency matches the characteristic frequency of our vortex system, it is, when the resonant absorption condition, $\omega\tau_1$$=$1, is fulfilled. As the driving frequency is fixed, we approach or de-tune from the resonant absorption condition by changing $\tau_1$ with temperature.
The non-monotonic temperature dependence of the phase shift reflects the non-trivial temperature dependencies of the different factors contributing in $\tau_1$.
\newline
\newline
In summary, we explored the microscopic linear response of a SC
ribbon to an external applied alternating magnetic field upon
decreasing temperature. A clear dichotomy between the ac-response of
the vortices and the screening currents was observed using the phase
sensitive character of the SSM technique. The observed out-of-phase
response of the vortices, which was absent for the reversible
screening currents, directly shows the local dissipation of the
vortex motion. As viscous losses can only account for a small
fraction of the observed dissipation at the used experimental
excitation frequency, the observed large phase-lag,
$\Theta$$=$-0.5rad, is explained by thermally activated vortex
hopping over pinning centers. As such a new and powerful tool is
introduced and employed allowing to investigate for the first time
vortex dynamics with single vortex resolution. This opens new and
exciting possibilities to study locally loss mechanisms in a variety
of interesting magnetic systems, including magnetic domain walls\cite{Kleemann, Saitoh}, etc.
\newline
\newline
We acknowledge the Methusalem funding by the Flemish government
and the Flemish Science Foundation, FWO-Vl, for financial support.

\end{document}